\documentclass[twocolumn, superscriptaddress, prl, longbibliography]{revtex4-2}
\usepackage{amsmath}
\usepackage{amsbsy}
\usepackage{amssymb}
\usepackage{graphicx}
\usepackage{color}
\usepackage{xcolor}
\usepackage{subfigure}
\usepackage{physics}
\usepackage{soul}
\usepackage{color}
\usepackage{bm}
\usepackage{array}
\usepackage{lipsum}
\usepackage[normalem]{ulem}
\usepackage{verbatim}
\usepackage{natbib}
\usepackage{nccmath}
\usepackage{algorithm2e}
\RestyleAlgo{ruled}
\usepackage[pdftex,colorlinks=true,
	pdfstartview = FitV,
	linkcolor    = linkcolor,
	citecolor    = linkcolor,
	urlcolor     = linkcolor,	
	hyperindex   = true,
	hyperfigures = false]{hyperref}
	
\definecolor{linkcolor}{rgb}{0,0,0.6}

\usepackage{cleveref}

\begin{document}

\title{Diffusive oscillators capture the pulsating states of deformable particles}

\author{Alessandro Manacorda}
\email{alessandro.manacorda@isc.cnr.it}
\affiliation{CNR Institute of Complex Systems, Uos Sapienza, Piazzale A. Moro 5, 00185 Rome, Italy}
\affiliation{Department of Physics and Materials Science, University of Luxembourg, L-1511 Luxembourg}
\author{\'Etienne Fodor}
\email{etienne.fodor@uni.lu}
\affiliation{Department of Physics and Materials Science, University of Luxembourg, L-1511 Luxembourg}

\date{\today}

\begin{abstract}
We study a model of diffusive oscillators whose internal states are subject to a periodic drive. These models are inspired by the dynamics of deformable particles with pulsating sizes, where repulsion leads to arrest the internal pulsation at high density. We reveal that, despite the absence of any repulsion between the diffusive oscillators, our model still captures the emergence of dynamical arrest. We demonstrate that arrest here stems from the discrete nature of internal states which enforces an effective energy landscape analogous to that of deformable particles. Moreover, we show that the competition between arrest and synchronisation promotes spiral waves reminiscent of the pulsating states of deformable particles. Using analytical coarse-graining, we derive and compare the collective dynamics of diffusive oscillators with that of deformable particles. This comparison leads to rationalize the emergence of spirals in terms of a rotational invariance at the coarse-grained level, and to elucidate the role of hydrodynamic fluctuations.
\end{abstract}

\maketitle

\textit{Introduction.}---Active matter encompasses systems where a constant energy injection at the particle level leads to collective dynamics far from equilibrium, as is the case for many chemical, biological, and robotic systems~\cite{Vicsek12pr, Marchetti13rmp, Bechinger16rmp, Fodor18physicaa}. In the last decades, most studies of active matter have focused on the role of self-propulsion, \emph{i.e.}~the ability of each particle to independently undergo directed motion. This paradigm has led to the theoretical understanding of several kinds of collective dynamics which have no counterpart in equilibrium systems~\cite{Cates2015, Chate2020}.

The energy injection in complex units need not result only in a self-propulsion mechanism. An important example is the case of deformable particles~\cite{Manning2023}. Indeed, complex aggregates such as macromolecules or living cells can change their shape due to internal activity~\cite{Zehnder15biophysj}, which leads to the spontaneous propagation of contraction waves in dense tissues. Such a wave propagation plays a crucial role in morphogenesis~\cite{Solon09cell, Martin09nature, Blanchard10development, Kruse2011, SerraPicamal12natphys, Heisenberg13cell}, uterine contraction~\cite{Xu15plos,Myers17wire}, and cardiac arrhytmogenesis~\cite{Karma13arcmp, Alonso16rpp}.

Vertex models are popular to capture the behavior of dense active systems such as biological tissues~\cite{Nagai01pmb}. They typically consider self-propulsion as the only active component and investigate how it affects the rigidity transition~\cite{Bi15natphys, Bi16prx}. Yet, when dense tissues behave like solids, it is questionable whether self-propulsion should be the key ingredient. Other models have considered dense assemblies of deformable particles~\cite{Tjhung17sm, Tjhung17pre, Brito18prx, Oyama19prr, Togashi19jcpb, Zhang23prl, Parisi23sr, Liu24njp, Ikeda24pre}, where energy injection occurs through the sustained oscillation of individual sizes.  With a Kuramoto-like synchronisation~\cite{Kuramoto1984,Acebron05rmp} of particle sizes, contraction waves spontaneously emerge~\cite{Togashi19jcpb, Zhang23prl, pineros2024}, which are reminiscent of those reported in biological systems~\cite{SerraPicamal12natphys, Karma13arcmp}.

\begin{figure}[b]
    \centering
    \includegraphics[width=.9\linewidth]{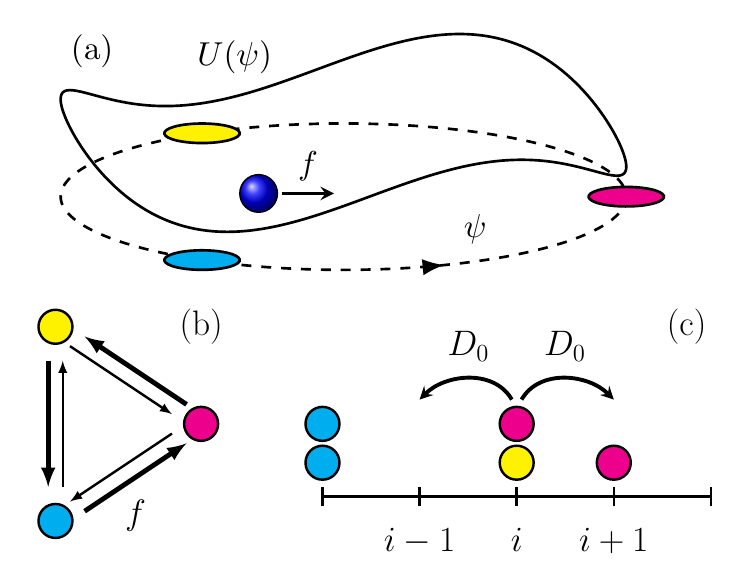}
    \caption{(a)~The collective dynamics of deformable particles maps into a continuous oscillator $\psi$ evolving in a potential $U$ and subject to a drive $f$~\cite[Sec.~S1.D]{SM}.  
    (b)~The internal dynamics of our discrete oscillator follows some transitions biased by the drive $f$ without any potential.
    (c)~Our oscillators freely diffuse at a rate $D_0$ without any volume exclusion.
    }
    \label{fig:schema}
\end{figure}

Contraction waves in deformable active particles stem from the competition between synchronisation and steric repulsion~\cite{Zhang23prl, pineros2024}. The former favors a global cycling of particle sizes. The latter favors some specific deformations, which is analogous to enforcing an external potential through which particle sizes are driven [Fig.~\ref{fig:schema}(a)]. At high density, local minima are deep enough to trap the particle size and thus promote an arrested state. Here, arrest refers to the hampering of size cycling, whereas it corresponds to the hampering of particle displacement within the rigidity transition of~\cite{Bi15natphys}.

Interestingly, there are alternative mechanisms for arrest and synchronisation beyond steric repulsion and Kuramoto-like interactions. This is a motivation for exploring whether these alternatives lead to a phenomenology similar to that of deformable active particles. Indeed, one may wonder whether the arrest-synchronisation competition is actually a generic scenario for pattern formation. If so, it encourages one to search for any hydrodynamic invariance and/or broken symmetry which might stand out as a hallmark of this competition.

\begin{figure}
    \centering
    \includegraphics[width=\linewidth]{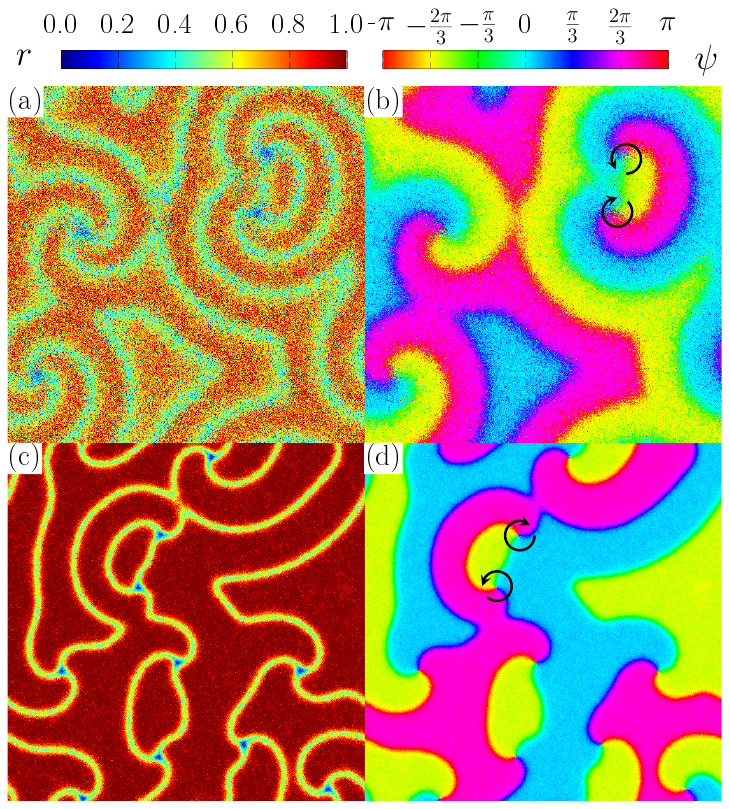}
    \caption{The amplitude $r$ and the phase $\psi$ of synchronization [Eq.~\eqref{eq:A}] reveal the coexistence of ordered domains in (a-b)~microscopic dynamics [Eq.~\eqref{eq:model}], and (c-d)~hydrodynamics [Eq.~\eqref{eq:hydro}]. 
    Rotating defects appear at the meeting point of domain interfaces, forming spiral waves with a threefold symmetry. (a)~$(L,f,\varepsilon,D_0,\rho_0)=(512,1.25,2.5,10^3,10)$, and (b)~$(L,f,\varepsilon,D,\rho_0) = (512,0.5,2.5,100,10)$.
    }
    \label{fig:snapshots_Lattice}
\end{figure}

In this Letter, we study a diffusive oscillator model (DOM) [Fig.~\ref{fig:schema}] featuring spiral waves rotating around defects [Fig.~\ref{fig:snapshots_Lattice}]. We reveal that the discreteness of internal states is key to arresting the dynamics, which, in combination with synchronisation and drive, generically yields wave propagation. Internal states can here be regarded as representing the local mimina of an external potential [Fig.~\ref{fig:schema}(a)], by analogy with the case of deformable particles~\cite{Zhang23prl, pineros2024}. Through analytical coarse-graining, we map our DOM into a specific form of the complex Ginzburg-Landau equation (CGLE)~\cite{Aranson02rmp} which breaks the {\it continuous} rotational invariance, while maintaining a {\it discrete} rotational invariance. We argue that such an invariance not only signals the emergence of arrest, but also constrains the features of the waves at the hydrodynamic and microscopic levels. Moreover, we reveal that hydrodynamic noise is essential to forming patterns, whereas density fluctuations are irrelevant.

Overall, our results demonstrate that the arrest-synchronisation competition, present in our DOM and in deformable particles~\cite{Zhang23prl, pineros2024}, is a generic route for forming patterns which are distinct from those of standard reaction-diffusion systems (RDS)~\cite{Murray2002, Murray2003, Odor04rmp}.


\textit{Diffusive oscillator model.}---We consider $N$ oscillators in $V=L^d$ sites of a hypercubic $d$-dimensional lattice, with global number density $\rho_0 = N/V$, without any excluded volume. Each site contains an arbitrarily large number of oscillators. Each oscillator has an internal state, labeled by a discrete index $a\in\{1,\ldots,q\}$, as a proxy to mimic the internal phase of deformable particles~\cite{Tjhung17sm, Tjhung17pre, Brito18prx, Oyama19prr, Togashi19jcpb, Zhang23prl, pineros2024}. The crucial difference is that such states now feature a {\it discrete} symmetry. In what follows, we focus on the case $q=3$, which is the minimum number of states to accommodate a current, and $d=2$. The system configuration $\{ n_{{\bf j},a} \}$ is then given by the number of oscillators for each state $a$ and site ${\bf j}$.

At every time step $dt/N$, an oscillator with state $a$ can either jump to a neighbouring site with probability $D_0 \, dt$, or switch to state $b$ with probability $W_{ba} \, dt$ [Figs.~\ref{fig:schema}(b,c)]:
\begin{equation}\label{eq:model}
	W_{ba} = \exp \left[ - f_{ba} + \frac{\varepsilon}{\rho_{\bf j}} \left(n_{{\bf j},b}-n_{{\bf j},a}\right) \right] .
\end{equation}
The first term in the exponent of Eq.~\eqref{eq:model} is defined by $f_{ab} = \pm f$ when $a-b= \pm 1 \mod 3$, and $f_{ab} = 0$ otherwise. This drive mimics the pulsation of deformable particles~\cite{Tjhung17sm, Tjhung17pre, Brito18prx, Oyama19prr, Togashi19jcpb, Zhang23prl, pineros2024}. Inspired by some active lattice dynamics~\cite{Thompson11jstatmech, Solon13prl, Solon15pre, Manacorda17prl, Manacorda2018, Kourbane18prl, Solon22prl}, the second term accounts for a synchronising Potts-like~\cite{Wu54rmp} interaction with strength $\varepsilon$, where $\rho_{\bf j} = \sum_a n_{{\bf j},a}$ is the local density at site $\bf j$. This term favors transition towards the state with the highest number of oscillators locally.

While interactions between oscillators are fully connected onsite, different sites exchange information only via diffusion. At small $D_0$, many transitions of internal states occur in between two rare jumps of sites. At large $D_0$, jumps are so frequent that now {\it all} oscillators are effectively interacting in between two state transitions. In both cases, patterns cannot emerge, and the system can be described solely in terms of the onsite dynamics~\cite{Herpich18prx, Herpich19pre}.

\textit{Breakdown of rotational invariance: Analogy with deformable particles.}---In the absence of diffusion ($D_0=0$), the evolution of the onsite occupation numbers $n_a$ can be studied in terms of the collective complex variable
\begin{equation}\label{eq:A_loc}
    A(t) = \frac{1}{\rho_0} \sum_{a=1}^3 e^{\frac{2\pi i}{3}a} n_a(t) \equiv r(t) \, e^{i \psi(t)} .
\end{equation}
Coarse-graining the microscopic dynamics, and expanding to the lowest orders in $A$, we get~\cite[Secs.~S1.A-B]{SM}
\begin{equation}\label{eq:dA}
    \dot A = c_1 A + c_2 {A^*}^2 + c_3 \vert A \vert^2 A \equiv {\cal L}(A)  ,
\end{equation}
where ${}^*$ refers to complex conjugation, and $(c_1, c_2, c_3)$ are complex coefficients fixed by $(f,\varepsilon)$~\cite[Sec.~S1.B]{SM}. In the regime of large $\rho_0$, fluctuations become irrelevant. The ${A^*}^2$ term in Eq.~\eqref{eq:dA} breaks the continuous symmetry $A \to A \,e^{i\phi}$, in contrast with the standard Stuart-Landau oscillator~\cite{PANTELEY2015645}. Yet, this term preserves the discrete symmetry $A \to A\, e^{\frac{2\pi i}{3} k}$ ($k$ integer). To the third order in $A$, Eq.~\eqref{eq:dA} actually contains all the terms compatible with this rotational symmetry.

\begin{figure}
    \centering
    \includegraphics[width=\columnwidth]{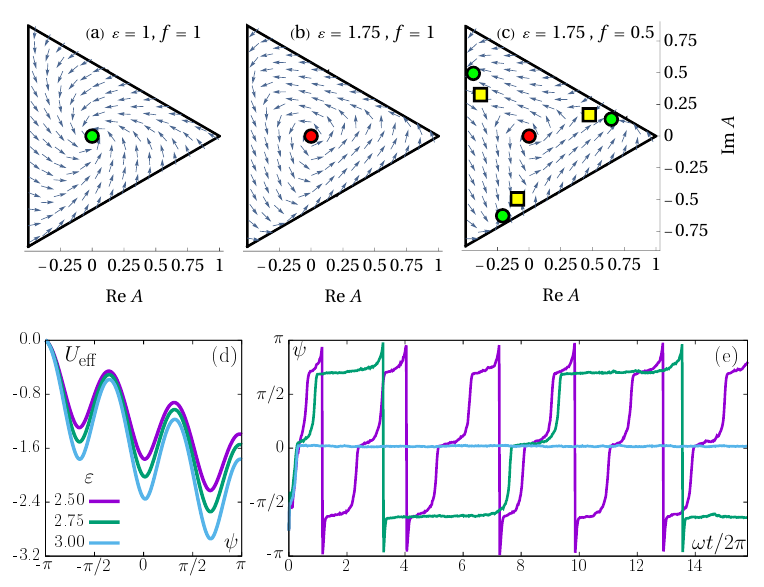}
    \caption{Stream plots of the collective variable $A$ [Eq.~\eqref{eq:dA}] for (a)~disorder, (b)~cycles, and (c)~arrest, with stable fixed points (green circles), unstable fixed points (red circles), and saddles (yellow squares).
    (d)~Effective landscape $U_{\rm eff}$ of the collective phase $\psi$ [Eq.~\eqref{eq:Ueff}] with $f=0.1$.
    (e)~Stochastic trajectories of $\psi$ measured in the onsite dynamics [Eq.~\eqref{eq:model}]. As $\varepsilon$ increases, phase oscillations slow down, and eventually reach arrest. $(N,f)=(10^3,2)$ and $\varepsilon$ as in (d). The natural pulsation reads $\omega=\sqrt3 \sinh f$~\cite[Sec.~S1.B]{SM}.
    }
    \label{fig:local}
\end{figure}

From the dynamics in Eq.~\eqref{eq:dA}, it follows that there are three stable phases~\cite{Herpich18prx, Herpich19pre}[Fig.~\ref{fig:local}]: (i)~a disordered phase at small $\varepsilon$, where oscillators are uniformly distributed in the three states, with symmetric fixed point $n_a = \rho_0/3$, and $|A|$ vanishes at large $\rho_0$; (ii)~a cycling phase at intermediate $\varepsilon$, where oscillators collectively undergo periodic transitions between states; (iii)~an arrested phase at large $\varepsilon$, with three fixed points invariant under the cyclic permutations $A\to A \,e^{\frac{2\pi i}{3} k}$.

The breakdown of the continuous symmetry $A \to A \,e^{i\phi}$ ensures the existence of arrest, while the discrete symmetry $A \to A\, e^{\frac{2\pi i}{3} k}$ enforces that the arrested phase is actually degenerate. Note that, in the microscopic model [Eq.~\eqref{eq:model}], transitions between states can be regarded as unimolecular reactions, in contrast with some models of multimolecular RDS~\cite{Reichenbach07nature, Reichenbach08jthbio}, and the total number of oscillators is conserved. These features lead to stabilising disorder at small $\varepsilon$, and ensure that arrest at large $\varepsilon$ is not an absorbing phase.

In the arrested phase, the discrete nature of the internal states enforces that the collective phase $\psi$ is subject to the effective landscape $U_{\rm eff}$~\cite[Sec.~S1.C]{SM}:
\begin{equation}\label{eq:Ueff}
    \dot{\psi} = - d U_{\rm eff}/ d\psi .
\end{equation}
Remarkably, $U_{\rm eff}$ features a series of local periodic minima whose depth increases with $\varepsilon$ [Fig.~\ref{fig:local}(d)], which can be rationalized in simple terms. At large $\varepsilon$, while continuous oscillators~\cite{Kuramoto1984} synchronously cycle without any cost, the collective cycling of our discrete oscillators [Eq.~\eqref{eq:model}] entails some periodic desynchronisations: the minima in $U_{\rm eff}$ describe the cost of such desynchronisations. As $\varepsilon$ increases, transitions between minima are less favored, so that our oscillators spend more time in a given state before cycling to the next one. Above a critical $\varepsilon$, this effect completely counteracts the drive, breaking down the periodicity of oscillations, and eventually stabilizing arrest [Fig.~\ref{fig:local}(e)].

Interestingly, a similar phase trapping has been reported in deformable particles~\cite{Togashi19jcpb, Zhang23prl, pineros2024}. Here, repulsion between particles is equivalent to an external potential constraining the deformation statistics [Fig.~\ref{fig:schema}(a)]. At the collective level, such a constraint can be recapitulated in terms of a landscape with periodic minima~\cite[Sec.~S1.D]{SM}, qualitatively analogous to $U_{\rm eff}$ [Fig.~\ref{fig:local}(d)]. In that respect, our DOM captures the same phenomenology, with identical collective states (disorder, arrest, cycles) as that of pulsating deformable particles.

In short, arrest emerges in our DOM solely due to the discreteness of internal states. As a result, our DOM entails a competition between arrest and synchronisation despite the absence of any repulsion, in contrast with deformable particles~\cite{Togashi19jcpb, Zhang23prl, pineros2024}. This competition opens the door to the emergence of dynamical patterns in spatially extended systems.


\textit{Spiral waves rotate around defects.}---In the presence of diffusion ($D_0>0$), the displacement of oscillators follows a free dynamics, independently of any interaction, so that the density profile is always homogeneous. Yet, the spatial distribution of the oscillator states may not remain homogeneous. Indeed, even when oscillators are synchronised on site, they might not be synchronised between sites, so that the system can potentially accommodate spatial instabilities.

To study the emergence of dynamical patterns, we introduce the {\it local} complex variable
\begin{equation}\label{eq:A}
	A_{\bf j}(t) = \frac{1}{\rho_{\bf j}(t)} \sum_{a=1}^3 e^{\frac{2\pi i}{3}a} n_{{\bf j},a}(t) \equiv r_{\bf j}(t) \, e^{i \psi_{\bf j}(t)} .
\end{equation}
When all oscillators are cycling in synchrony, the amplitude $r_{\bf j}\approx 1$ and the phase $\psi_{\bf j}$ are homogeneous. The period of $\psi_{\bf j}$ increases with $\varepsilon$, and eventually diverges. Before diverging, it undergoes large temporal fluctuations, which may desynchronise nearby sites, thus promoting spatial fluctuations of $\psi_{\bf j}$. At large $D_0$, spatial fluctuations are suppressed by the rapid displacement of oscillators in the system. Instead, at moderate $D_0$, such fluctuations can potentially build up into a large-scale instability.

\begin{figure}
  \centering
  \includegraphics[width=\columnwidth]{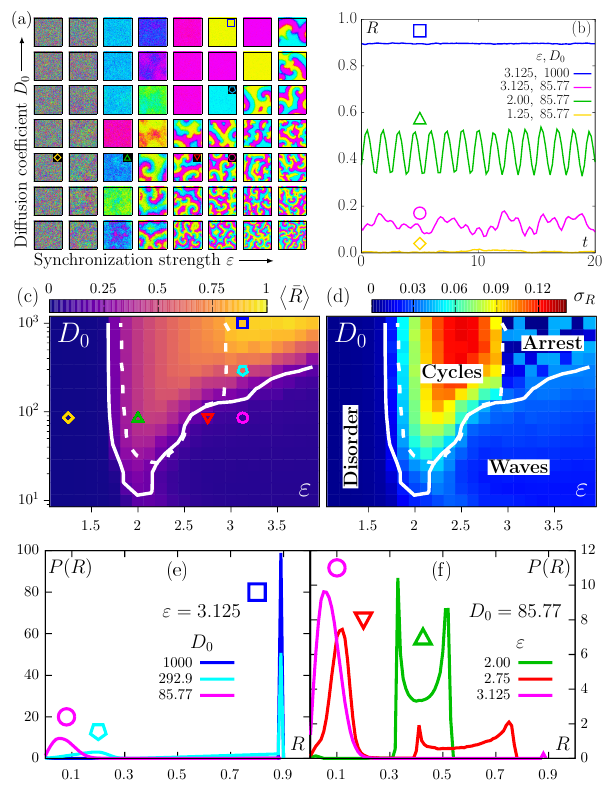}
  \caption{(a)~Snapshots of the local phase $\psi_j$ [Eq.~\eqref{eq:A}, color code in Fig.~\ref{fig:snapshots_Lattice}].
  (b)~Trajectories of the synchronisation parameter $R(t)$ [Eq.~\eqref{eq:order}].
  (c)~The averaged parameter $\langle\bar{R}\rangle$ and (d)~its standard deviation $\sigma_R$ [Eq.~\eqref{eq:stat}] lead to guidelines delineating boundaries between four phases. The solid and dashed lines correspond respectively to $\langle\bar{R}\rangle=0.2$ and $\sigma_R=0.045$. Symbols refer to the parameter values taken for panels~(a-b) and~(e-d). (e,f)~Distribution of $R$ for various phases (1024 realizations). The arrested-wave and cycling-wave transitions both display metastability. $(L,\rho_0)=(128,10)$, and $f=\varepsilon/2$.
  }
  \label{fig:PhaseDiagram_Lattice}
\end{figure}

Instabilities can lead to the spatial coexistence of three cycling domains where $r_{\bf j}\equiv r_{{\bf j},c}\approx 1$. Interfaces between domains have a finite width where $r_{{\bf j},c}>r_{\bf j}>0$. The meeting points of interfaces are given by defects where $r_{\bf j}\approx 0$. Since all domains cycle at the same frequency, defects effectively rotate, thus forming spiral waves with a threefold symmetry [Figs.~\ref{fig:snapshots_Lattice}(a-b)]. Defects connected by the same interface rotate in opposite directions, and can annihilate by pairs when colliding. Higher $D_0$ increases the domain sizes and interface widths, thus reducing the number of defects. Higher $\varepsilon$ reduces the interface widths and increases $r_{{\bf j},c}$ [Fig.~\ref{fig:PhaseDiagram_Lattice}(a)].

Remarkably, we do not observe any states with turbulent waves, in contrast with the case of continuously deformable particles~\cite{Zhang23prl}. Such a turbulence stems in~\cite{Zhang23prl} from excitations of the homogeneous arrested phase which promote some localized, aperiodic cyclings. The discrete symmetry of our DOM, which entails three arrested states, prevents such events by trapping the phase before it completes one cycle. Therefore, waves in our DOM spontaneously organize into steady spirals with threefold symmetry. Note that the merging of defects can actually also stabilize planar waves. Besides, when initially ordered, the system can also accommodate circular waves without any defect~\cite[Sec.~S2.A]{SM}.


\textit{Phase boundaries and transitions.}---In addition to the phase with waves, we also observe the emergence of three homogeneous phases (disorder, cycles, arrest) analogous to the onsite case. To quantitatively distinguish these phases, we introduce the synchronisation parameter
\begin{equation}\label{eq:order}
	R(t) = \frac1{N} \bigg| \sum_{{\bf j},a} e^{\frac{2\pi i}{3} a} \, n_{{\bf j},a}(t) \bigg| .
\end{equation}
The trajectories of $R(t)$ allow us to identify four scenarios [Figs.~\ref{fig:PhaseDiagram_Lattice}(a,b)]: (i)~a disordered phase at small $\varepsilon$ [$R(t)\approx 0$, yellow line]; (ii)~a cycling phase at intermediate $\varepsilon$ and large $D_0$ [oscillating $R(t)$, green line]; (iii)~an arrested phase at large $\varepsilon$ and large $D_0$ [$R(t) \lesssim 1$, blue line]; (iv)~spiral waves at large $\varepsilon$ and intermediate $D_0$ [$R(t)$ strongly fluctuates, pink line].

To delineate phase boundaries, we consider the time-averaged $\bar R$ and the variance $\sigma^2_R$ of $R(t)$ [Eq.~\eqref{eq:order}]:
\begin{equation}\label{eq:stat}
	\bar{R} = \int^{t_0+t}_{t_0} \frac{ R(u) \,\mathrm{d} u}{t} ,
	\;
	\sigma^2_R = \int^{t_0+t}_{t_0} \big\langle \big[R(u)-\bar{R}\big]^2 \big\rangle \frac{\mathrm{d} u}{t} ,
\end{equation}
where $\langle\cdot\rangle$ is an average over realizations. $\langle\bar R\rangle$ is clearly smaller in the disordered and wave phases (without global synchronisation) than in the cycling and arrested phases (with global synchronisation) [Fig.~\ref{fig:PhaseDiagram_Lattice}(c)]. The variance $\sigma_R$ is higher in the cycling phase than in all others [Fig.~\ref{fig:PhaseDiagram_Lattice}(d)]. Indeed, $R(t)$ strongly oscillates in this phase [Fig.~\ref{fig:PhaseDiagram_Lattice}(b)] due to periodic desynchronisation. Overall, $\langle\bar R\rangle$ and $\sigma_R$ yield the phase boundaries in Figs.~\ref{fig:PhaseDiagram_Lattice}(c-d).

We examine how the distribution $P(R)$ varies across transitions. Deep in the wave and arrested phases [resp. pink and dark blue lines], $P(R)$ has a single peak, respectively at small and large $R$. Deep in the cycling phase [green line], $P(R)$ is nonzero for a finite domain of $R$ and peaks at its boundaries, due to the oscillations of $R(t)$. Going from waves to arrest, $P(R)$ becomes bimodal, thus signaling metastability [Fig.~\ref{fig:PhaseDiagram_Lattice}(e)]. Similarly, going from waves to cycling, $P(R)$ is now nonzero in two separate domains, due to the coexistence between the two dynamical states [Fig.~\ref{fig:PhaseDiagram_Lattice}(f)]. In short, both the arrest-wave and cycling-wave transitions feature a metastable regime.


\textit{The role of hydrodynamic fluctuations.}--Although spiral waves have already been reported in many RDS, one may wonder how our patterns [Figs.~\ref{fig:snapshots_Lattice}(a-b)] actually differ from standard instabilities present, for instance, in the CGLE~\cite{Aranson02rmp}. To address this question, we coarse-grain our DOM in terms of $A_{\bf j}(t) \to {\cal A}({\bf x},t)$ and $\rho_{\bf j}(t)\to \rho({\bf x},t)$ in the continuum limit, and expand to the lowest orders in ${\cal A}$~\cite[Sec.~S1.B]{SM}:
\begin{equation}\label{eq:hydro}
    \partial_t {\cal A} = D \nabla^2 {\cal A} + {\cal L}({\cal A}) + \Lambda ,
\end{equation}
where $\cal L$ is defined in Eq.~\eqref{eq:dA}, and $D \propto D_0$ is the macroscopic diffusion coefficient. Neglecting the fluctuations of the coarse-grained density $\rho({\bf x},t)$, we deduce that it obeys the simple diffusion equation ($\partial_t\rho = D\nabla^2\rho$) independently of $\cal{A}$, and relaxes to the homogeneous profile $\rho=\rho_0$. Thus, we approximate $\Lambda$ as an additive zero-mean Gaussian white noise with correlations proportional to $1/\rho_0$~\cite[Sec.~S1.B]{SM}.

The degeneracy of the arrested state directly affects the shape of the hydrodynamic patterns. Indeed, Eq.~\eqref{eq:hydro} entails an instability promoting the spatial coexistence of cycling domains. Rotating defects with threefold symmetry spontaneously form where interfaces meet [Figs.~\ref{fig:snapshots_Lattice}(c-d)], yielding the same spiral waves as in our DOM [Figs.~\ref{fig:snapshots_Lattice}(a-b)]. Remarkably, in the absence of noise ($\Lambda=0$), the homogeneous states are always stable~\cite[Sec.~S1.B]{SM}. In other words, while density fluctuations can be safely neglected, the fluctuations in the hydrodynamics of $\cal A$ are essential to capture patterns, as in~\cite{Zhang23prl}.

In short, our coarse-graining shows that the hydrodynamics of our DOM is distinct from the standard CGLE~\cite{Aranson02rmp}, so that our DOM clearly differs from standard RDS. In practice, adding to the standard CGLE the lowest-order nonlinearities, compatible with the discrete symmetry, suffices to reproduce the specific shape of spirals observed in our DOM.


\textit{Discussion.}---Our DOM with discrete symmetry entails spiral waves stemming from the competition between arrest and synchronisation [Fig.~\ref{fig:snapshots_Lattice}]. The key idea is that discrete states enforce an effective landscape [Fig.~\ref{fig:local}(d)] equivalent to the case of deformable particles with repulsive interactions~\cite[Sec.~S1.D]{SM}\cite{Zhang23prl, pineros2024}. Discreteness of states here suffices to promote arrest, thus providing a mechanism distinct from that at play in~\cite{Zhang23prl, pineros2024}. Therefore, our results show that the arrest-synchronisation scenario for pattern formation extends to a broad class of models with discrete symmetry. Remarkably, waves are shaped by the discrete rotational invariance at the hydrodynamic level, which distinguishes them from the patterns of other RDS~\cite{Murray2002, Murray2003, Odor04rmp}. Besides, the fluctuations of the complex field are crucial to yield patterns, whereas density fluctuations are irrelevant.

Our work paves the way to examining the interplay between discrete symmetry and pattern formation for arbitrary $q$. Indeed, we expect that the three homogeneous phases (disorder, cycles, arrest) are robust beyond $q=3$. To study patterns, one can use a top-down approach postulating the hydrodynamics by identifying the terms which obey the discrete symmetry ${\cal A}\to{\cal A}e^{\frac{2\pi i}{q} k}$. As in our coarse-graining~\cite[Sec.~S1]{SM}, one can also derive the hydrodynamic coefficients in terms of the microscopic parameters. It is tempting to speculate that, depending on the parity of $q$, such a hydrodynamic study could lead to identifying generic properties of defect dynamics~\cite{Shankar22natrevphys, Mahault22natcomm}, and potential connections with the topology of other active models with phase trapping~\cite{Solon22prl, avni2023, avni2024, Noguchi24jcp, Noguchi24njp}.

The discrete symmetry of our DOM seems to preclude a defect turbulence, at variance with the standard CGLE~\cite{Aranson02rmp} and its recent generalization~\cite{Zhang23prl}. To capture such a turbulence, one could introduce energy levels which maintain the discrete nature of our DOM while breaking its symmetry. The arrested phase would no longer be degenerate, thus opening the door to local excitations nucleating defects from a homogeneous phase, as reported in~\cite{Zhang23prl, pineros2024}. In this context, it would be interesting to explore how density fluctuations affect the defect nucleation at the hydrodynamic and microscopic levels~\cite{banerjee2024}. Finally, our DOM can be straightforwardly adapted to account for thermodynamic consistency~\cite{Agranov_2024}, allowing one to study the energetics of the corresponding patterns~\cite{meibohm2024, meibohm2024bis, avanzini2024}.


\acknowledgments{We acknowledge fruitful discussions with L. K. Davis, M. Esposito, J. Meihbom, and Y. Zhang. This project has received funding from the European Union’s Horizon Europe research and innovation programme under the Marie Sk\l{}odowska-Curie grant agreement No 101056825 (NewGenActive), and from the Luxembourg National Research Fund (FNR), grant reference 14389168. A.M. acknowledges financial support from the project MOCA funded by MUR PRIN2022 grant No. 2022HNW5YL}

\bibliography{pulsating}

\end{document}


\title{Supplemental material:\\Diffusive oscillators capture the pulsating states of deformable particles}

\author{Alessandro Manacorda}
\email{alessandro.manacorda@isc.cnr.it}
\affiliation{CNR Institute of Complex Systems, Uos Sapienza, Piazzale A. Moro 5, 00185 Rome, Italy}
\affiliation{Department of Physics and Materials Science, University of Luxembourg, L-1511 Luxembourg}
\author{\'Etienne Fodor}
\email{etienne.fodor@uni.lu}
\affiliation{Department of Physics and Materials Science, University of Luxembourg, L-1511 Luxembourg}

\date{\today}

\maketitle


\section{Coarse-graining from microscopics to hydrodynamics}

\subsection{Path-integral representation of stochastic dynamics}

The microscopic dynamics of occupation numbers in discrete time reads
\begin{equation}\label{eq:lattice}
n_{\ii,a,s+1} = n_{\ii,a,s} + J_{\ii,a,s} ,
\end{equation}
being $J_{\ii,a,s}=0,\pm 1$ the increment of species $a$ at site $\ii$ and time step $s$.  Following the procedure in~\cite{Thompson11jstatmech} we derive the hydrodynamic equations from the equation above by writing the probability of a single trajectory $\{n_{\ii,a,s}\}$ for $a=0,1,2$, $\ii\in\L$ and 
$s=0,\ldots,T/\de t$, which reads
\begin{equation}\label{eq:prob-traj}
\mathbb{P} \argc{\arga{n_{\ii,a,s}}} = \moy{\prod_{\ii,s,a} \delta \argp{n_{\ii,a,s+1} - 
n_{\ii,a,s} - J_{\ii,a,s}}}_\JJJ \ ,
\end{equation}
being $\moy{\cdot}_\JJJ$ the average over the possible realizations of $\JJJ=\arga{J_{\ii,a,s}}$. Expanding the delta 
functions in their integral representation we have
\begin{equation}\label{eq:prob-traj-exp}
\mathbb{P} \argc{\arga{n_{\ii,a,s}}} = \int \prod_{\ii,s,a} \left( \de \hn_{\ii,a,s} \, 
e^{\hn_{\ii,a,s} \argp{n_{\ii,a,s+1} - n_{\ii,a,s}}} \moy{ e^{- \hn_{\ii,a,s} J_{\ii,a,s}}}_\JJJ 
\right) \ .
\end{equation}
The last term must be computed looking at the microscopic probabilities defined in 
the main text. At a given time $p$, we have the following options: 
\begin{itemize}
    \item a jump from site $\ii$ to site $\ii+\s \eee_\m$, being $\s=\pm1$ and $\eee_\m$ the unit 
    vector in the $\m$-th direction, occurring with probability $D_0\de t$, yielding $J_{\ii,a,s}=-1$ and 
    $J_{\ii+\s \eee_\m,a,s}=+1$;
    \item a transition from state $a$ to state $b$ on the site $\ii$, occurring with probability 
    $W_{ba}\argp{\bar{n}_{\ii,a,s}}$, yielding $J_{\ii,a,s}=-1$ and $J_{\ii,b,s}=+1$;
    \item a null event, occurring with probability $1-\sum_{\ii,a} \argp{ 2d D_0 + \sum_{b} W_{ba}} \de t $.
\end{itemize}
In the definition above, we indicate with $\bar{n}_{\ii,a,s}$ the occupation numbers involved in the transition rates, which can in principle refer to neighboring sites (not necessarily nearest-neighbors), as in~\cite{Thompson11jstatmech}. In this work transition rates are computed at zero-range, \ie taking into account only particles in the same site, but we keep the derivation general to facilitate further work. We can therefore write the average over $\JJJ$ as
\begin{equation}\label{eq:exp-av}
\begin{split}
\prod_{\ii,a} \moy{e^{- \hn_{\ii,a,s} J_{\ii,a,s}}}_\JJJ &\simeq 1 + \sum_{\ii,a} \sum_{\m,\s} 
\argp{ e^{\hn_{\ii,a,s} - \hn_{\ii+\s \eee_\m,a,s}} - 1} n_{\ii,a,s} D_0 \de t + \sum_{\ii,a} 
\sum_{b} \argp{ e^{\hn_{\ii,a,s} - \hn_{\ii,b,s}} - 1} n_{\ii,a,s} W_{ba}(\bar \nn_{\ii,s}) \de t \\
&\simeq \prod_{\ii,a} \exp \left\{ \sum_\m \argc{ \argp{ e^{\hn_{\ii,a,s} - \hn_{\ii+ \eee_\m,a,s}} - 1} 
n_{\ii,a,s} + \argp{e^{\hn_{\ii+\eee_\m,a,s} - \hn_{\ii,a,s}} - 1} n_{\ii+\eee_\m,a,s} } D_0 \de t \right. 
\\
&\qquad \qquad \quad \left. + \sum_{b} \argp{ e^{\hn_{\ii,a,s} - \hn_{\ii,b,s}} - 1} 
n_{\ii,a,s} W_{ba}(\bar \nn_{\ii,s}) \de t \right\} .
\end{split}
\end{equation}
We inject the last result in Eq.~\eqref{eq:prob-traj-exp}, and consider the continous-time limit through the substitutions:
\begin{equation}
n_{\ii,a,s+1}-n_{\ii,a,s} \to \dot{n}_{\ii,a} \de t \ ; \qquad \sum^{T/\de t}_{s=1} \de t \to \int^T_0 
\de t \ ; \qquad \prod_s \de \hn_{\ii,a,s} \to \DD \hn_{\ii,a} \ ,
\end{equation}
so that the probability of a trajectory reads
\begin{equation}\label{eq:prob-S}
\mathbb{P} \argc{\arga{n_{\ii,a}(t)}} = \int \prod_{\ii,a} \DD \hn_{\ii,a} \, e^{-S\argc{\nn,\hat{\nn}}} \ , 
\end{equation}
with
\begin{equation}\label{eq:action}
\begin{split}
S &= - \int_0^T \de t \sum_{\ii,a} \arga{ \hn_{\ii,a} \dot{n}_{\ii,a} + D_0 \sum_\m \argc{ \argp{ 
e^{\hn_{\ii,a} - \hn_{\ii+ \eee_\m,a}} - 1} n_{\ii,a} + \argp{e^{\hn_{\ii+\eee_\m,a} - \hn_{\ii,a}} - 1} 
n_{\ii+\eee_l,a} } \right.
\\
&\qquad \left. + \sum_{b} \argp{ e^{\hn_{\ii,a} - \hn_{\ii,b}} - 1} n_{\ii,a} 
W_{ba}(\bar \nn_\ii) } \ .
\end{split}
\end{equation}
The continuous-space limit can be taken by introducing the lattice spacing $\l$ and substituting
\begin{equation}
n_{\ii,a} \to \l^d \r_a(\xx) \ ; \qquad \hn_{\ii,a} \to  \hr_a(\xx) \ ; \qquad  \sum_{\ii \in \L} \to 
\int_{V=[0:\ell]^d} \de \xx \, \l^{-d} \ ; \qquad \nabla_{\ii,\m} \to \l \dpar{x_\m} + \frac12 \l^2 \frac{\partial^2}{\partial x_\m^2}
\ ,
\end{equation}
being $\ell=L\l$ the lattice length and $\nabla_{\ii,\m} f_\ii \equiv f_{\ii+\eee_\m}-f_\ii$ the 
$l$-th coordinate of the discrete gradient. After summing over the coordinates $\m$, the leading order 
in the Taylor expansion of the action reads
\begin{equation}\label{eq:S0}
S = S_0 + \OO(\l) = - \int^T_0 \de t \int_V \de \xx \, \sum_a \argc{ \hr_a \dot\r_a + D \argp{ \vert \nabla \hr_a 
\vert^2 \r_a + \nabla \r_a \cdot \nabla \hr_a } + \sum_{b} \argp{ e^{\hr_a - \hr_b} - 1} \r_a 
W_{ba}(\bar \br ) } + \OO(\l) \ ,
\end{equation}
having defined $D = D_0 \l^2 = \OO(1)$ in the hydrodynamic limit $\l\to 0 \ , \: \ell=\OO(1)$.
To proceed further, we truncate the exponential of response fields at second order, 
which corresponds to an approximation of Gaussian noise for the fluctuating hydrodynamics. The action 
then reads
\begin{equation}\label{eq:S0-trunc}
\begin{split}
S_0 &= - \int^T_0 \de t \int_V \de \xx \, \sum_a \arga{ \hr_a \dot\r_a + D \argp{ \vert \nabla \hr_a 
\vert^2 \r_a + \nabla \r_a \cdot \nabla \hr_a } + \sum_{b} \argc{ \hr_a - \hr_b + \frac12 
\argp{\hr_a - \hr_b}^2} \r_a W_{ba}(\bar \br) } \\ 
&= - \int^T_0 \de t \int_V \de \xx \, \sum_a \left\{ \hr_a \dot\r_a + D  \argp{ \vert \nabla \hr_a 
\vert^2 \r_a + \nabla \r_a \cdot \nabla \hr_a } + \hr_a \sum_{b} \argc{ W_{ba}(\bar \br) \r_a - 
W_{ab}(\bar \br) \r_b } \right. \\
& \qquad \qquad \qquad \qquad \qquad \left. + \frac12 \sum_{b(>a)} \argp{\hr_a - \hr_b}^2 \argc{ 
W_{ba}(\bar \br) \r_a + W_{ab}(\bar \br) \r_b } \right\} \ .
\end{split}
\end{equation}
The quadratic terms in the response fields can be removed through a Hubbard-Stratonovich transformation introducing the auxiliary fields $\bx_a(\xx,t)$ (vectorial in space) and $\h_{ab}(\xx,t)$, so that (after integration by parts)
\begin{equation}\label{eq:S0-HubbardStratonovich}
\begin{split}
\mathbb{P} \argc{\br(\xx,t)} &= \frac1{Z} \int \DD [\hr_a,\x_a,\h_{ab}] \, e^{-S_0[\br,\hat{\br},\bx,\bh]} , \\
S_0 &= - \int^T_0 \de t \int_V \de \xx \, \sum_a \left\{ \hr_a \dot\r_a - \frac12 \x^2_a - \hr_a \argc{ 
\nabla \cdot \argp{\sqrt{2 D \r_a} \, \bx_a} + D \nabla^2 \r_a } \right. \\
& \quad \quad \quad \left. + \hr_a \sum_{b} \argc{ W_{ba}(\bar \br) \r_a - W_{ab}(\bar \br) \r_b }
+ \sum_{b(>a)} \argc{ -\frac12 \h^2_{ab} + \sqrt{ W_{ba}(\bar \br) \r_a + W_{ab}(\bar \br) \r_b } \argp{\hr_a - \hr_b} \h_{ab} } \right\} \ .
\end{split}
\end{equation}
In the last line, the noise terms $\h_{ab}$ are white and Gaussian; however, if we want to integrate over $\DD \hr_a$ in the first equation, we need to collect the terms separately. We thus define for $a<b$ the antisymmetric matrix $\D_{ab}$ and the noise field $\eta_a$ as
\begin{equation}
\D_{ab} (\bar\br) = \sqrt{ W_{ba}(\bar \br) \r_a + W_{ab}(\bar \br) \r_b } = -\D_{ba}(\bar\br) \ ,
\quad
\h_a = \sum_{b(\neq a)} \D_{ab} \h_{ba} \,
\end{equation}
so that
\begin{equation}\label{eq:noise_tr}
\begin{split}
	\sum_{a<b} \D_{ab} \argp{\hr_a - \hr_b} \h_{ab} &= \sum_a \h_a \hr_a \ ,
	\\
  \moy{\x_{a\m}({\bf x},t) \x_{b\n}({\bf x}',t')} &= \d_{ab} \d_{\m\n} \delta({\bf x}-{\bf x}') \delta(t-t') \ ,
  \\
   \moy{\h_a({\bf x},t) \h_b({\bf x}',t')} &= \bigg[ \d_{ab} \sum_{c} \D^2_{ac}(\bar\br({\bf x},t)) - (1-\d_{ab}) \D^2_{ab}(\bar\br({\bf x},t)) \bigg] \delta({\bf x}-{\bf x}') \delta(t-t') \ .
\end{split}
\end{equation}
We rewrite the action further as
\begin{equation}
\begin{split}
S_0 &= - \int^T_0 \de t \int_V \de \xx \, \sum_a \left\{ \hr_a \argc{ \dot \r_a -
\nabla \cdot \argp{\sqrt{2 D \r_a} \, \bx_a} - D \nabla^2 \r_a - \sum_{b} \argp{ W_{ab}(\bar \br) \r_b - 
W_{ba}(\bar \br) \r_a } - \h_a } \right.
\\
&\left.\qquad - \frac12 \x^2_a  - \sum_{b(>a)} \frac12 \h^2_{ab} \right\} \ .
\end{split}
\end{equation}
Integrating over the response fields $\hr_a$, we obtain
\begin{equation}
\begin{split}
\mathbb{P} \argc{\br(\xx,t)} &\propto \int \DD[\x_a,\h_{ab}] e^{-\int \de t \de x \sum_a (\x^2_a/2 + 
\sum_b \h^2_{ab}/2) }
\\
&\quad\times \prod_a \delta \argp{\dot \r_a - \nabla \argp{\sqrt{2 D \r_a} \, \bx_a} - D \nabla^2 \r_a 
- \sum_{b} \argp{ W_{ab}(\bar \br) \r_b - W_{ba}(\bar \br) \r_a } - \h_a}  \ ,
\end{split}
\end{equation}
yielding the fluctuating hydrodynamic (FHD) equations:
\begin{equation}\label{eq:HD-rho}
  \dot \r_a = - \nabla \cdot \JJJ_a + T_a \ ,
  \quad
  \JJJ_a = - D \nabla \r_a + \sqrt{2 D \r_a} \ \bx_a \ ,
  \quad
  T_a = \sum_{b} \argc{ W_{ab}(\bar \br) \r_b - W_{ba}(\bar \br) \r_a } + \h_a \ .
\end{equation}
We operate the linear change of variables leading us to the density field $\r$ and to the complex order parameter $\AA$
\begin{equation}
\r (\xx,t) = \sum_a \r_a(\xx,t)  \ , \quad \AA (\xx,t) = \frac1{\r (\xx,t) }\sum_a e^{\frac{2\p i}{q}a} \r_a(\xx,t)  \ .
\end{equation}
The FHD equations become then
\begin{equation}\label{eq:HD-rho-A}
\begin{split}
  \partial_t \r &= - \nabla \cdot \JJJ \ , \quad
  \JJJ = - D \nabla \r + \sqrt{2 D \r} \ \bx \ , \quad
  \bx = \sum_a \sqrt{\r_a/\r} \ \bx_a \ ,\\
  \partial_t (\r \AA) &= - \nabla \cdot \JJJ_A + T_A \ , \quad
  \JJJ_A = - D \nabla (\r \AA) + \sqrt{2 D \r} \ \bx_A \ , \quad
  \bx_A = \sum_a e^{\frac{2\pi i}{q}a} \sqrt{\r_a/\r} \ \bx_a  \ , \\
  &\moy{\argp{\x_\m,\x_{A,\m},\x^*_{A,\m}}^\dagger(\xx,t) \argp{\x_\n,\x_{A,\n},\x^*_{A,\n}}(\xx',t')} = 
  \begin{pmatrix}
  1 & \AA & \AA^* \\
  \AA^* & 1 & \AA \\
  \AA & \AA^* & 1
  \end{pmatrix} \d_{\m\n} \delta(\xx-\xx')\delta(t-t') \ , \\
  T_A &= \sum_{ab} e^{\frac{2\pi i}{q}a} \argc{ W_{ab}(\bar \br) \r_b - W_{ba}(\bar \br) \r_a } + \h_A \ , \\
  \h_A &= \sum_a e^{\frac{2\pi i}{q}a} \h_a = \sum_{ab} e^{\frac{2\pi i}{q}a} \D_{ab} \h_{ba} \ .
\end{split}
\end{equation}
The $\D_{ab}$ terms explicitly read
\begin{equation}\label{eq:Deltas}
\begin{split}
    \D^2_{01} &= \frac23 \r \argc{  \cosh\Omega_1 + e^{-i\p/3} \AA     
    \cosh\argp{\Omega_1+i\p/3} + e^{i\p/3} \AA^* \cosh\argp{\Omega_1-i\p/3} } \ , \quad     \Omega_1 = f - \frac{\ee}{\sqrt3} \argp{ e^{i\p/6}\AA+e^{-i\p/6}\AA^*} \ ,\\
    \D^2_{02} &= \frac23 \r \argc{ \cosh\Omega_2 + e^{i\p/3} \AA 
    \cosh\argp{\Omega_2+i\p/3} + e^{-i\p/3} \AA^* \cosh\argp{\Omega_2-i\p/3} } \ 
    , \quad \Omega_2 = f + \frac{\ee}{\sqrt3} 
    \argp{ e^{-i\p/6}\AA+e^{i\p/6}\AA^*} \ , \\
    \D^2_{12} &= \frac23 \r \argc{ \cosh\Omega_3 - i \AA 
    \sinh\argp{\Omega_3-i\p/6} + i \AA^* \sinh\argp{\Omega_3+i\p/6} } \ , 
    \quad \Omega_3 = f + \frac{\ee}{\sqrt3} 
    i \argp{\AA-\AA^*} \ . \\
\end{split}
\end{equation}


\subsection{Expansion around the disordered state}

\begin{figure}
    \centering
    \includegraphics[width=\textwidth]{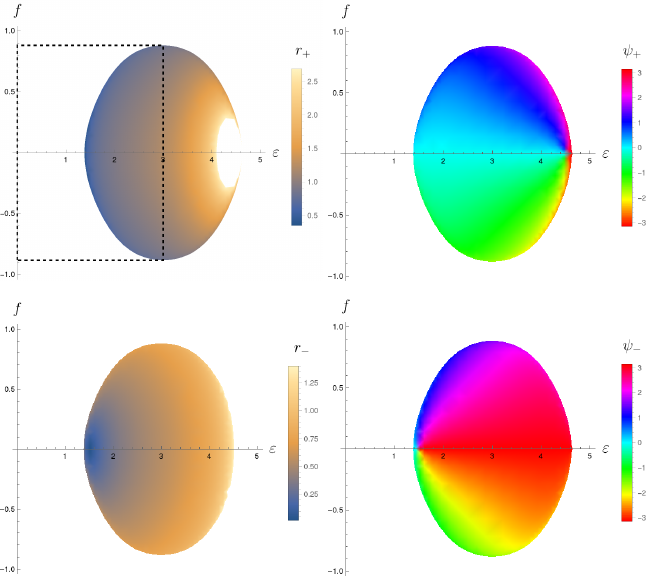}
    \caption{The two ordered fixed points $(r_\pm,\psi_\pm)$ of Eq.~\eqref{eq:R-psi-hydro} in the synchronisation-frequency plane. The dashed rectangle in top left panel shows the physical domain of validity of the hydrodynamics.}
    \label{fig:SM1}
\end{figure}

We expand the hydrodyamic equations~\eqref{eq:HD-rho-A} around the fixed point $\AA=0$, assuming $\vert \AA \vert \ll 1$. Then, since the evolution of $\r$ does not depend on $\AA$, the density relaxes to an homogeneous profile with Gaussian fluctuations around $\r = \r_0$. We then neglect current fluctuations (which can be interpreted as a large-size limit $L\gg1$) and replace $\r(x,t)$ with a constant value $\r_0$  in the equations for $\AA$. We expand the transition rates in $T_A$ up to the third order in $(\AA,\AA^*)$, following a Landau-Ginzburg scheme. We therefore get
\begin{equation}\label{eq:dA-expansion}
\frac1{\r_0} T_A \simeq c_1 \AA + c_2 {\AA^*}^2 + c_3 \vert \AA \vert^2 \AA + \frac1{\r_0} \h_A \ ,
\end{equation}
where
\begin{equation}\label{eq:c}
c_1 = (2\ee-3) \cosh f + i \sqrt3 \sinh f \ , \quad
c_2 = \ee \left( \cosh f + i \frac{\ee-3}{\sqrt3} \sinh f \right) \ , \quad c_3 = \frac{\ee^2}6 \left[ (2\ee-9) \cosh f + i \sqrt3 \sinh f \right] \ .
\end{equation}
This result sets a first condition for our expansion: one must have $\ee<9/2$ to avoid the divergence of the complex field. The noise can also be expanded noting that, when $\AA=\AA^*=0$, one has from Eq.~\eqref{eq:Deltas}
\begin{equation}
\begin{split}
&\moy{\h_A(\xx,t) \h^*_A(\xx',t')} = \Big[ 6 \r_0 \cosh f + \OO(\AA) \Big] \delta(\xx-\xx')\delta(t-t') \ ,
\\
&\moy{\h_A(\xx,t)\h_A(\xx',t')} = \moy{\h^*_A(\xx,t)\h^*_A(\xx',t')} = \OO(\AA) \delta(\xx-\xx')\delta(t-t') \ .
\end{split}
\end{equation}
Keeping only the first-order additive terms in the noise, one finds the hydrodynamics in 
Eq.~(6) of the main text:
\begin{equation}\label{eq:hydro-SM}
\begin{split}
    \partial_t \AA &= D \nabla^2 \AA + c_1 \AA + c_2 {\AA^*}^2 + c_3 \vert \AA \vert^2 \AA + \L \ , \\    
    \moy{\L(\xx,t) \L^*(\xx',t')} &= \frac6{\r_0} \cosh f \, \d\argp{\xx-\xx'} \d \argp{t-t'} \ , \\
    \moy{\L(\xx,t) \L(\xx',t')} &=\moy{\L^*(\xx,t) \L^*(\xx',t')} = 0 \ .
\end{split}
\end{equation}
The CGLE global invariance $\mathcal A \to \mathcal A e^{i\phi}$ is explicitly broken by the term ${\mathcal A^*}^2$, yet the $C_3$ discrete symmetry $\mathcal A \to \mathcal A e^{\frac{2\p i}3 k}$ holds for an arbitrary integer $k$. The noiseless dynamics in the phase-amplitude coordinates $\mathcal A = r e^{i\psi}$ reads
\begin{equation}\label{eq:R-psi-hydro}
\begin{split}
\partial_t r &= D\left( \nabla^2 r - r \vert \nabla \psi \vert^2 \right) + a_1 r + \vert c_2 \vert \cos \left( \phi_2 - 3\psi \right) r^2 + a_3 r^3 \ , \\
r \partial_t \psi &= D \left( r \nabla^2 \psi + 2 \nabla r \cdot \nabla \psi \right) + b_1 r + \vert c_2 \vert \sin \left( \phi_2 - 3\psi \right) r^2 + b_3 r^3 \ ,
\end{split}
\end{equation}
where $c_j = a_j+i b_j = \vert c_j \vert e^{i \phi_j}$. Since $\dot\psi = b_1$ when $\ee=0$, we define a natural oscillation frequency $\om=b_1=\sqrt3 \sinh f$. The non-trivial homogeneous fixed points $(r_\pm,\psi_\pm)$ of Eq.~\eqref{eq:R-psi-hydro} are given by
\begin{equation}\label{eq:FP}
\vert c_3 \vert^2 r^4_\pm + \left( c_1 \bar{c}_3 + \bar{c}_1 c_3 - \vert c_2 \vert^2 \right) r^2_\pm + \vert c_1 \vert^2 = 0 \ ,
\quad
\tan \left( \phi_2 - 3\psi_\pm \right) = \frac{b_1 + b_3 r^2_\pm}{a_1+a_3 r^2_\pm} \ .
\end{equation}
We remark that the angle $\psi$ is multiplied by $3$ in both equations; this means that any fixed point $\psi_\pm$ entails two symmetric fixed points with $(r_\pm', \psi_\pm') = (r_\pm, \psi_\pm \pm 2\p/3)$, as a consequence of the discrete symmetry of the system. We follow two particular solutions in the $(-\p,\p]$ range, acknowledging that the symmetric solutions always exist. The fixed points $(r_\pm,\psi_\pm)$ in the $(\ee,f)$ plane are shown in Fig.~\ref{fig:SM1}.

We now focus on linear stability. The stability of the homogeneous disordered state $\AA=0$ is determined by the sign of $a_1 = \Re c_1$. Indeed, to linear order in absence of noise, one has
\begin{equation}
    \partial_t \AA = D \nabla^2 \AA + c_1 \AA \quad \Rightarrow \quad 
    \partial_t \hat\AA = \left( c_1 - D q^2 \right) \hat\AA \ ,
\end{equation}
moving to Fourier space, \ie $\hat \AA (\qq,t) = \int \de \xx e^{-i\qq\cdot\xx} \AA(\xx,t)$, and $q^2 = \vert\qq\vert^2$. This gives $\l(q) = a_1 - Dq^2 + i b_1$: the homogeneous disordered state is stable if $\Re \l(q)<0$ for all $q$, which yields $a_1 <0$ \ie $\ee<3/2$. The linearized hydrodynamic equations around the non-trivial homogeneous fixed points is more involved. In polar coordinates, the dynamics of the perturbation $\d r = r - r_\pm$ and $\d\psi = \psi-\psi_\pm$ is given by
\begin{equation}\label{eq:linearized}
    \begin{split}
        \partial_t \d r &= \left[D \nabla^2 + a_1 + 2 \vert c_2 \vert r_\pm \cos(\phi_2-3\psi_\pm) + 3 a_3 r^2_\pm \right] \d r + 3 \vert c_2 \vert r^2_\pm \sin (\phi_2 - 3\psi_\pm) \d\psi \ , \\
        \partial_t \d\psi &= \left[ 2 \vert c_2 \vert \sin(\phi_2 - 3\psi_\pm) + 2 b_3 r_\pm \right] r_\pm \d r + \left[ D \nabla^2 - 3 \vert c_2 \vert r_\pm^2 \cos(\phi_2 - 3\psi_\pm) \right] \d\psi \ .
    \end{split}
\end{equation}
The corresponding eigenvalue problem in Fourier space then reads
\begin{equation}
    \frac{\de}{\de t} 
    \begin{pmatrix}
        \d\hat r \\
        \d\hat\psi 
    \end{pmatrix}
    = \left( M_0 - D q^2 \, I\right)
    \begin{pmatrix}
        \d\hat r \\
        \d\hat\psi 
    \end{pmatrix} \ .
\end{equation}
 $M_0$ is the matrix of the global modes defined by the coefficients of $\d r$ and $\d\psi$ in Eq.~\eqref{eq:linearized}, excluding the Laplacian which enters through $-D q^2$. The eigenvalues $\l(q)$ can be simply obtained by the eigenvalues $\l_0$ of the collective mode $q=0$, \ie $\l(q) = \l_0 - Dq^2$. The linear stability of the homogeneous fixed points is then determined by the eigenvalues of $M_0$. In particular, one needs $\lambda_0<0$ in both eigenspaces to ensure stability. Solving the eigenproblem numerically in the $(f,\ee)$ plane, one finds
 that $(r_+,\psi_+)$ is stable within almost all the existence region, except for a small area close to $(f,\ee)=(0,9/2)$ which is beyond the physical domain of our equations. Conversely, the fixed point $(r_-,\psi_-)$ is always unstable. Therefore, the dynamics in the physical region always converges to $(r_\pm,\psi_\pm)$, when it exists.

These fixed points represent the ordered, arrested solution of the dynamics. We therefore observe three regions:
\begin{enumerate}
\item A region of stability of the disordered phase $\ee<3/2$.
\item A region of existence and stability of the ordered phase, shown in Fig.~\ref{fig:SM1}, leading to arrest.
\item A region where no fixed point is stable, leading to limit cycles since $\dot \psi \neq 0$.
\end{enumerate}
The region in Fig.~\ref{fig:SM1} defines a critical line $f_c(\ee)$, and the ordered solution exists for $f < f_c(\ee)$. Above $\ee=3$ one has $\de f_c / \de \ee<0$: this behavior is not physical as we know that higher synchronisation should always promote order. We therefore take $\ee=3$ as an upper bound of the validity domain of our approximation. At the same time, the ordered fixed points exist only below the drive threshold $f_c(\ee=3)=\tanh^{-1}(1/\sqrt2) \simeq 0.88$, above which order cannot be achieved for any synchronisation strength. We furthermore remark that, in the same region $\ee>3$, there are solutions with $r>1$ which have to be considered spurious as well - as from our definition in the main one always needs to have $0\leq r \leq1$. Therefore, the physical domain of validity of the third-order expansion is $\vert f \vert < f_c(3)$ and $0<\ee<3$. This region is shown by the dashed black line in Fig.~\ref{fig:SM1}.


\subsection{Effective potential for diffusive oscillator model}\label{sec:dom}

\begin{figure}[!t]
    \centering
    \includegraphics[width=\linewidth]{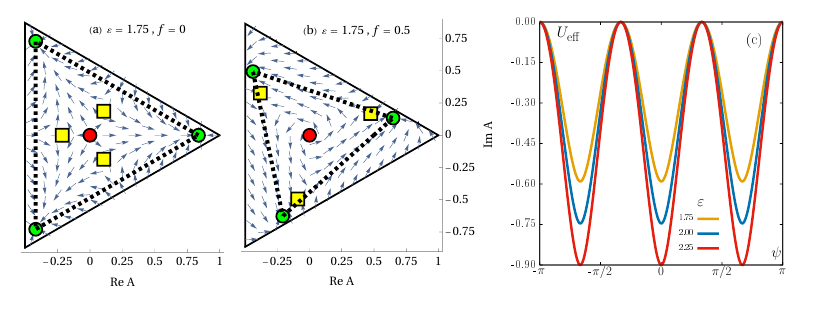}
    \caption{Fixed points in the ordered phase (green circles), as in Fig.~3(c) of the main text, for (a)~$f=0$, and (b)~$f>0$. The dashed black lines connect the fixed points through the amplitude-phase relation in Eq.~\eqref{eq:r-psi}. (c)~Effective potential $U_{\rm eff}(\psi)$ for $f=0$ analytically computed in Eq.~\eqref{eq:Ueff0}.}
    \label{fig:SM2}
\end{figure}

We now demonstrate how to define an effective potential for the phase $\psi$ from the dynamics in Eq.~\eqref{eq:R-psi-hydro}. We first eliminate the variable $r$ from the evolution of $\psi$ by approximating it with the line connecting two ordered fixed points in the arrested phase. Namely, in the complex plane $r(\cos\psi,\sin\psi)\equiv(x,y)$, this line is given by
\begin{equation}
    \frac{y-y_1}{y_2-y_1} = \frac{x-x_1}{x_2-x_1} \ ,
    \quad (x_1,y_1) = r_+(\cos\psi_+,\sin\psi_+) \ ,
    \quad
    (x_2,y_2) = r_+(\cos(\psi_+ + 2\p / 3),\sin(\psi_+ + 2\p / 3)) \ ,
\end{equation}
where $(r_+,\psi_+)$ is given in Eq.~\eqref{eq:FP}. Note that $\psi_+$ can be any of the three ordered fixed points. Moving to polar coordinates, we have
\begin{equation}\label{eq:eff}
    \frac{r \cos\psi - r_+ \cos\psi_+}{r_+ \cos \left( \psi_+ + 2\p/3 \right) - r_+ \cos\psi_+} = \frac{r \sin\psi - r_+ \sin\psi_+}{r_+ \sin \left( \psi_+ + 2\p/3 \right) - r_+ \sin\psi_+} \ .
\end{equation}
The amplitude-phase relation of the fixed points [Eq.~\eqref{eq:FP}] can be substituted in Eq.~\eqref{eq:eff}, and replacing $\psi - \psi_+$ with its modulo $2\pi/3$ makes the relation $r(\psi)$ valid for any choice of the initial $\psi_+$, yielding
\begin{equation}\label{eq:r-psi}
    r(\psi) = \frac{r_+(\varepsilon,f)}{2 \cos\left( \d\psi - \p/3 \right)} \ , \quad  \d\psi = \left[ \psi - \psi_+(\varepsilon,f) \right] \mod 2\p/3 \ .
\end{equation}
The amplitude $r$ is represented as a function of the phase $\psi$, with black dashed lines connecting the fixed points, in Figs.~\ref{fig:SM2}(a-b). The corresponding phase dynamics [Eq.~\eqref{eq:R-psi-hydro}] then reads
\begin{equation}
    \dot\psi = b_1 + \vert c_2 \vert \sin(\phi_2 - 3\psi) r(\psi) + b_3 r^2(\psi) \equiv -\partial_\psi U_{\rm eff}(\psi,\varepsilon,f) \ .
\end{equation}
The effective potential $U_{\rm eff}$ is shown in Fig.~3(d) of the main text. As expected, increasing $\varepsilon$ yields higher energy barriers in $\psi$, thus making the fixed points more stable and increasing the escape time from a stationary solution. When $f=0$, the expression of $U_{\rm eff}$ greatly simplifies:
\begin{equation}\label{eq:Ueff0}
    U_{\rm eff}(\psi,\varepsilon,f=0) = -\frac{\varepsilon \, r_+(\varepsilon,f=0)}{2} \left[ \log \cos \left( \d\psi - \frac{\p}3 \right) + 2 \sin^2 \left( \d\psi - \frac{\p}3 \right) \right] \ .
\end{equation}
The effective potential in the ordered phase is shown in Fig.~\ref{fig:SM2}(c). In particular, we confirm that the barrier height
\begin{equation}
    \D U(\ee) = U_{\rm eff}(\psi=\p/3,\ee) - U_{\rm eff}(\psi=0,\ee) = \frac{3-\log 4}{8(9-2 \ee)} \sqrt{ 3 \sqrt3 \sqrt{8\ee(6-\ee)-51} + 12 \left( \ee (6-\ee) -6 \right)}
\end{equation}
increases with $\ee$.

\subsection{Effective potential for pulsating deformable particles}

Pulsating deformable particles have been studied in a model with positions $\rr_i$ in and internal phases $\th_i$~\cite{Zhang23prl}. The latter determine the particles' radii $\s_i$ through the relation $\s(\th) = \s_0 (1+\l \sin\th)/(1+\l)$, where $\s_0$ is the maximum radius. The many-body dynamics reads
\begin{equation}\label{eq:PAM}
    \dot \rr_i = - \m_r \sum_{j \wedge i} \partial_{\rr_i} U_{ij} + \sqrt{2\m_r T} \, \h_r \ ,
    \quad
    \dot \th_i = \om + \sum_{j \wedge i}\Big[ \ee \sin(\th_j - \th_i) - \mu_\th \partial_{\th_i} U_{ij} \Big] + \sqrt{2 \m_\th T} \, \h_\th \ ,
\end{equation}
where $j\wedge i$ refers to a summation over the neighbors of particle $i$. The noise terms $(\h_r,\h_\th)$ are uncorrelated and Gaussian, with zero mean and unit variance. The potential $U = \sum_{i, j<i} U_{ij} = \sum_{i, j<i} V ( \vert \rr_i - \rr_j \vert/(\s_i+\s_j))$ implements some repulsive, short-ranged interactions (e.g., choosing $V$ as a WCA potential), so that the phase dynamics arrests at high density $\r_0 = N/V$: it corresponds to a mechanical balance between the drive $\om$ and the repulsive force $-\mu_\th \sum_j \partial_{\th_i} U_{ij}$.

As shown in Ref.~\cite{Zhang23prl}, coarse-graining the dynamics in Eq.~\eqref{eq:PAM} proceeds in two steps:
\begin{itemize}
    \item Neglecting the repulsive interactions in the position dynamics ($\partial_{\rr_i} U_{ij} \to 0$). This amounts to assuming that the density field relaxes to a uniform profile.
    \item Approximating the repulsive force in the phase dynamics as $\partial_{\th_i} U_{ij} \approx c \partial_{\th_i} \ph$, in terms of the packing fraction $\ph = \frac{\p}{V} \sum_i \s(\th_i)^2$, where $c = \partial_\varphi U$ is assumed constant.
\end{itemize}
The onsite dynamics of the collective complex variable $A=\frac 1 N \sum_j e^{i\theta_j}$ follows as~\cite{Zhang23prl}
\begin{equation}\label{eq:hydro-PAM}
    \dot A = c_1 A + c_3 \vert A \vert^2 A + c_4  A^2 + c_4 \vert A \vert^2 + c_6 A^3 - c_6^* \vert A \vert^2 A^* \ ,
\end{equation}
where
\begin{equation}
    c_1 = \frac{\ee \r_0}2 - D_\th + i \om \ , \quad c_3 = - \frac{\ee(2 D_\th + i\om)}{4(4D_\th+\om^2)}  \ , \quad c_4 = -i \frac{c\mu_\theta \lambda \sigma(0)^2}{2V} \ , \quad c_6 = - \frac{c\ee\mu_\theta \lambda^2 \sigma(0)^2(2 D_\th + i\om)}{8V(4D_\th+\om^2)}  \ .
\end{equation}
The dynamics in Eq.~\eqref{eq:hydro-PAM} can be written in a phase-amplitude representation as
\begin{equation}
    \dot r = a_1 r + a_3 r^3 \ ,
    \quad
    r \dot\psi = b_1 r + 2 b_4 r^2 \cos \psi + \left[ b_3 + 2 \vert c_6 \vert \sin \left( \phi_6 + 2\psi \right) \right] r^3 \ ,
\end{equation}
where $c_j = a_j + i b_j = \vert c_j \vert e^{i \phi_j}$. The amplitude dynamics does not depend on the phase $\phi$, so that the amplitude $r$ relaxes towards its stable fixed point $r = \sqrt{-a_1/a_3}$ in the ordered phase. Then, one can substitute $r$ in the dynamics of $\psi$, yielding
\begin{equation}
    \dot\psi = - \partial_\psi U_{\rm PAM}(\psi) \ , \quad U_{\rm PAM}(\psi) = \frac14 \left( \frac{\ee\r_0}{D_\th} - 6 \right) \om\psi + \frac{c\sqrt{4 D^2_\th + \om^2}}{\ee} \left[ \sqrt{\frac{\ee\r_0}{2D_\th}-1} - \frac{\l}4 \left( \frac{\ee\r_0}{2D_\th}-1 \right) \sin \left( \phi_6 + \psi\right) \right] \sin \psi \ .
\end{equation}
The potential $U_{\rm PAM}$ has a single minimum for $\psi\sin(0,2\pi)$, whose depth increases with $\ee$ as for the case of the diffusive oscillator model [Sec.~\ref{sec:dom}]. In practice, changing the specific radius-phase relation $\s(\th)$ can lead to multiple minima.



\section{Numerical methods}

\subsection{Simulation of Microscopic lattice dynamics}

Simulations are performed on a square lattice in $d=2$ dimensions. Every particle is labeled with a state $a\in\{1,\ldots,q\}$, with $q=3$ from now on. The microscopic configuration of the system is given by the positions ${\bf i}_n \in \{1,L\}^2$ and states $\sigma_n$ of each particle $n\in\{1,\ldots,N\}$ at a given discrete time $t = s {\rm d}t/N$. The system is invariant under relabelling of particles, which preserves the occupation numbers
\begin{equation}
n_{{\bf j},a} = \sum^N_{n=1} \delta_{{\bf i}_n,{\bf j}} \delta_{\s_n,a} \ ,
\end{equation}
accounting for the number of particles at site $\jj$ with state $a$. The microscopic configuration at discrete time $s$ is then univocally determined by the occupation numbers. The lattice size is $V = L^2$ and the number of particles is $N = \rho_0 V$.  Simulations are performed with $L=128$ in the phase diagram shown in Fig.~2(c-d) of the main text, and  $L=512$ in Fig.~1(c-d); $\rho_0=10$ in both cases.

At every time step, a particle $n$ is chosen with uniform random distribution $p_n = 1/N$, and a uniform number $\xi \in [0,1]$ is drawn to decide whether (i)~the particle jumps to a next-neighbor site with probability $D_0 \,{\rm d}t$, (ii)~the particle switches from state $a$ to state $b$ with probability $W_{ba} \, {\rm d}t$, or (iii)~none of the above with probability $1- \left( 2dD_0 +\sum_{b\neq a} W_{ba} \right) \de t$. To ensure normalization, the last term must be positive; the time step $\de t $ is then taken as
\begin{equation}
	\de t = \left[ 2d D_0+ 
	e^{f+\varepsilon} + e^{- f} \right]^{-1} \lesssim \bigg( 2dD_0 + \max_{\{ n\}} \sum_{b\neq a} W_{ba} \bigg)^{-1}  ,
\end{equation}
assuming $\varepsilon>0$ as usual. After each move, the time is advanced by ${\rm d}t/N$, so that each particle attempts a move on average once at each $t \to t+ {\rm d}t$~\cite{Kourbane18prl}. Simulations have been performed via the ULHPC cluster and parallelised with the use of GNU parallel~\cite{Tange2018}. Snapshots in Fig.~1(c-d) and phase diagrams of Fig.~2(c-d) of the main text are realised with 128 realizations starting from initially disordered configurations, \ie every particle's position and state are independent and drawn from a uniform distribution. Realizations are performed until $t_{\rm max}=200$ in our system, and statistical measures (averages and standard deviations) are performed in the time window $t_0<t<t_{\rm max}$, where $t_0=150$, consisting of 256 measures. The phase boundaries in Fig.~2(c-d) of the main text are drawn via a linear interpolation of the raw data, thus locating the isosurfaces of constant $\left\langle \bar R \right\rangle$ or $\sigma_R$. The dynamics shown in the snapshots in Fig.~1(c-d) of the main text can be found in the video~\texttt{Lattice.mp4}.

\subsubsection{Transition from cycles to arrest in the onsite dynamics ($D_0=0$)}

We report in Fig.~\ref{fig:SM3} the transition from cycles to arrest for the onsite dynamics mentioned in the main text. The arrested fixed points are given by the asymmetric stationary solutions of the master equation
\begin{equation}
\dot n_a = \sum_{b (\neq a)} \left[ W_{ab} (\nn) \, n_b - W_{ba}(\nn) \, n_a 
\right] \ .
\end{equation}
At small $\varepsilon$, the phase oscillates with a periodic behavior [purple line in Fig.~3(e)], whereas the phase arrests at large $\varepsilon$ [cyan line in Fig.~3(e)]. In between these regimes, fluctuations can lead to escape from arrest, so that the phase $\psi$ spend a large random time close to a fixed point, before a random event allows it to continue cycling [green line in Fig.~3(e)].

\subsubsection{Wave propagation with ordered initial conditions}

\begin{figure}
    \centering
    \includegraphics[width=\textwidth]{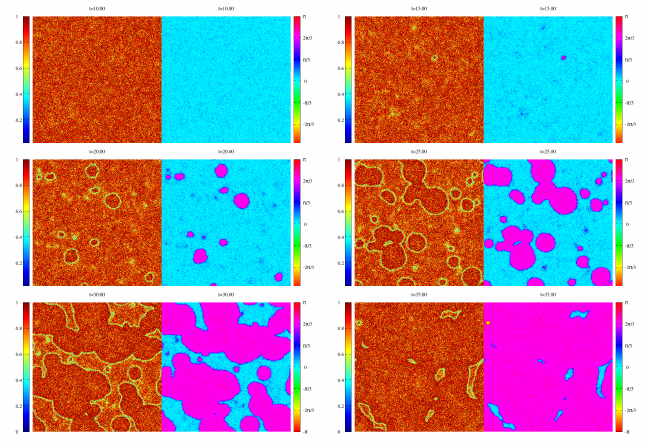}
    \caption{Nucleation and expansion of stable phases destabilizing the arrested  homogeneous state. Phase-amplitude plots, respectively $r$ (left panels) and $\psi$ (right panels) for each snapshot, at times $10<t<35$. Color codes as in Fig.~2 of the main text Parameters: $(L,f,\ee,D_0,\r_0) = (512,1.5,3,10^2,10)$. }
    \label{fig:SM3}
\end{figure}

The spiral wave propagation described in the main text takes place when the system is quenched from a disordered initial condition. Instead, when the initial condition is ordered, the homogeneous ordered state is destabilized by nucleation of droplets. This behavior is shown in Fig.~\ref{fig:SM3}. At $t=10$, the system is ordered and homogeneous with $\psi(\xx) \simeq 0$. At $t=15$, small droplets with $\psi \simeq 2\p/3$ are visible. At $t=20$ more droplets have formed and the previous ones are expanding. At $t=25,30$, droplets merge and the state $\psi \simeq 2\p/3$ invades the system. At $t=35$, the state $\psi \simeq 0$ has almost disappeared; a droplet with $\psi \simeq -2\p/3$ is visible in the top left region, starting again the nucleation destabilizing the homogeneous profile $\psi \simeq 2\p/3$.

In a similar model, spiral waves have also been observed starting from initially ordered conditions~\cite{Noguchi24njp}. In this paper, the authors refer to the state where stable phases nucleate in an arrested background [Fig.~\ref{fig:SM3}] as ``homogeneous cycling''. Instead, in our work, we identify homogeneous cycling as the regime where the phase profile remains homogeneous and undergoes cycling, for which the hydrodynamic equations have no homogeneous fixed points.

\subsubsection{Interface width and domain size}

\begin{figure}
    \centering
    \includegraphics[width=\textwidth]{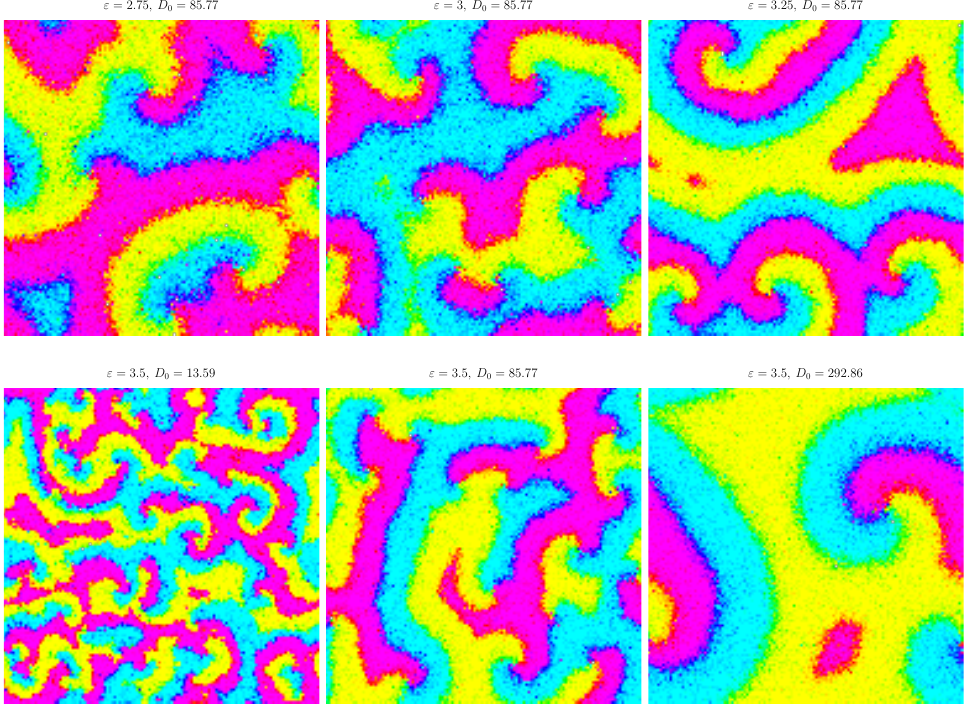}
    \caption{Snapshots of the phase field in the wave regime to illustrate the effect of synchronisation and diffusion. Top panels: synchronisation increases from left to right.
    Interfaces between domains become more regular as fluctuations are suppressed by strong coupling. Bottom panel: diffusion increases from left to right. Higher diffusion leads to larger domains. Parameters: $(L,f,\r_0)=(128,\ee/2,10)$.}
    \label{fig:SM4}
\end{figure}

During wave propagation, higher synchronisation sharpens the interfaces between domains, and higher diffusion increases the domains' size. This effect is visible in Fig.~\ref{fig:SM4}. Both effects have an energetic motivation: interfaces correspond to intermediate synchronisation at the saddle points of  the underlying energetic landscape, and are therefore penalized by higher synchronisation. Diffusion allows particles to interact over higher length scales, leading to larger domains. Diffusion thus entails a characteristic length scale $l = \sqrt{D \t}$, related to characteristic times of the dynamics such as pulsation $\t_f = \cosh f$ or nucleation time $\t_{\rm nuc}$. When this length scale is comparable with the system size ($l \sim L$) the system becomes homogeneous.

\subsection{Simulation of hydrodynamic equations}
 
Eqs.~\eqref{eq:hydro-SM} are simulated with a spectral method to treat the (linear) diffusion term. We simulate the equations in the range $(x,y)\in[0:\ell]^2$ and $t>0$, and perform a linear discretisation of space and time, namely $(x,y)=(i,j)\D x$ and $t=n\D t$, with $\D x=L/\ell=1$ and $\D t=10^{-3}$, being now $L$ the number of spatial nodes. Simulations start from a disordered configuration, taking $A_{ij,n=0} = A_0 (\x_R+i\x_I)$, being $\x_{R/I}$ two Gaussian unitary random numbers. The discrete evolution $t \to t+\D t$ occurs in the following steps:

\begin{enumerate}
 \item the local evolution is performed with a fourth-order Runge-Kutta 
 scheme on the deterministic part, and the noise is added 
 within the Euler-Maruyama scheme~\cite{Gardiner2009}. The Runge-Kutta scheme, 
 even if not needed when noise is present, makes the discretisation robust also 
 in the zero-noise limit~\cite{Liu19pre}. The local fields are evolved as
 $A_{ij,n}\to A_{ij,n+1/2}$;
 \item the evolved field is Fourier transformed and we get $\hat A_{hk,n+1/2} = 
 {\rm FFT}[A_{n+1/2}]$;
 \item the modes $(h,k) \in \{ -L/2,\ldots,L/2-1 \}^2$ are evolved locally taking into account only the 
 diffusive term with a trapezoidal integration over the interval $[t,t+\D t]$, 
 \ie 
\begin{equation}
    \hat A_{hk,n+1} = \frac{1-q^2 D \D t/2}{1+q^2 D \D t/2} \hat A_{hk,n+1/2}
\end{equation}
being $q^2 = \left(\frac{2\p}{\ell}\right)^2 (h^2+k^2)$;
 \item the Fourier modes are antitransformed and we obtain the evolved complex field $A_{ij,n+1} =  {\rm FFT}^{-1}[\hat A_{n+1}]$.
\end{enumerate}
The dynamics shown in the snapshots in Fig.~4(d-e) of the main text can be found in the video \texttt{Hydro.mp4}.

\bibliographystyle{apsrev4-2}
\bibliography{pulsating}